\documentstyle[aps,amsfonts,amssymb,multicol,prl]{revtex}
\input{psfig.sty}

\newcommand{\ep}{\varepsilon}
\renewcommand{\k}{{\bf k}}
\newcommand{\R}{{\bf R}}
\newcommand{\x}{{\bf x}}

\renewcommand{\Im}{\textrm{Im}\,}
\renewcommand{\Re}{\textrm{Re}\,}

\begin{document}

\title{
Theory of the Fano Resonance in the STM Tunneling Density of States due to a
Single Kondo Impurity}
\author{O. \'Ujs\'aghy$^{a,b}$, J. Kroha$^d$, L. Szunyogh$^a$ and 
A. Zawadowski$^{a,b,c}$}
\address{$^a$Department of Physics and $^b$Research Group of the Hungarian
Academy of Sciences\\ Technical University of Budapest,
H-1521 Budapest, Hungary}
\address{$^c$Solid State and Optical Research Institute of the Hungarian
Academy of Sciences,  POB 49, H-1525 Budapest, Hungary}
\address{$^d$Institut f\"ur Theorie der Kondensierten Materie, University of Karlsruhe, POB 6980,
D-76128 Karlsruhe, Germany}
\date{\today}

\draft
\maketitle

\begin{abstract}
The conduction electron density of states nearby single magnetic impurities,
as measured recently by scanning tunneling microscopy (STM), 
is calculated, taking into account tunneling into conduction electron 
states only.
The Kondo effect induces a narrow Fano resonance in the conduction
electron density of states, while scattering off the d-level
generates a weakly energy dependent Friedel oscillation.
The line shape varies with the distance between STM tip
and impurity, in qualitative agreement with experiments, but
is very sensitive to details of the band structure.
For a Co impurity the experimentally observed width and shift of the 
Kondo resonance are in accordance with those obtained from a combination of
band structure and strongly correlated calculations.
\end{abstract}

\pacs{PACS numbers: 72.15.Qm,72.10.Fk,61.16.Ch}

\begin{multicols}{2}

\begin{figure}
\centerline{\psfig{figure=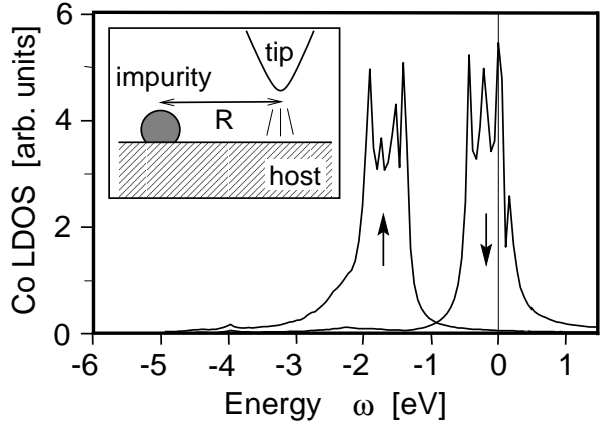,width=8truecm}}
\narrowtext
\vspace*{0.5cm}
    \caption{LSDA result for the 
    LDOS of a Co impurity on a Au(111) surface, showing 
    majority ($\uparrow$) and minority ($\downarrow$) states.
    The inset shows the experimental setup schematically.}
    \label{fig:szk}
\end{figure}

Recently, several groups have demonstrated using scanning tunneling
microscopy \cite{Li,Madhavan,Manoharan} that a magnetic Kondo impurity
adsorbed on the surface of a normal metal causes a narrow, resonance-like
structure in the electronic surface density of states (DOS), whose asymmetric
line shape resembles that of a Fano resonance \cite{Fano}. 
The experiments were performed with single Ce atoms on Ag \cite{Li} as 
well as with single Co atoms on Au \cite{Madhavan} and Cu \cite{Manoharan} 
surfaces by measuring the I-V
characteristics of the tunneling current through the tip of a 
scanning tunneling microscope (STM) placed close to the surface 
and at a small distance $R$ from the magnetic atom 
(see Fig.~\ref{fig:szk}(inset)). 

Although the Kondo resonance, formed in the local conduction electron 
density of states (LDOS) at the Fermi energy $\ep _F$ 
due to resonant spin flip scattering, has been known for a long 
time \cite{Mezei1}, the precise line shape    
was not studied earlier because of the limited spatial resolution
available in experiments. Wide tunnel junctions, 
which were proposed as 
measurement devices \cite{Mezei2}, probe the averaged DOS rather than the LDOS.
Such experiments exhibited a giant zero bias resistance peak \cite{Berman,Cooper}
or a weak conductance peak, the latter induced by assisted tunneling through
impurities in the tunnel barrier \cite{Appelbaum,Wyatt}. 

In the present Letter we present a detailed  microscopic study of the 
Kondo line shape as measured by STM in the vicinity of a single magnetic
ion. We assume that the STM current is predominantly due to tunneling into the
conduction LDOS, i.e. we neglect direct tunneling into the d- or f-level
of a Co or Ce ion. This assumption is justified, because the d- or f-level
is localized deeply in the atomic core, and is sufficient to explain
the experimental findings as seen below.
When a discrete (single-particle) level is coupled to the 
conduction electron sea (bare DOS $\rho_0$)
via a hybridization $V$, there is a twofold effect: 
(1) the discrete level is broadened and (2) the continuous conduction
LDOS becomes modified. The resulting line shape in the LDOS is called
Fano resonance, in reminiscence of the first study of this problem in the
context of atomic physics \cite{Fano}. Here we generalize this problem to 
the interacting case, i.e. when the discrete level arises from a 
many-body effect, like the Kondo resonance. It is shown that the many-body
correlation effects and the consecutive Fano line shape in the conduction LDOS
may be understood in separate steps, thus greatly simplifying the
theoretical treatment as compared to other studies 
\cite{Madhavan,Schiller}.

\begin{figure}[htbp]
\centerline{\psfig{figure=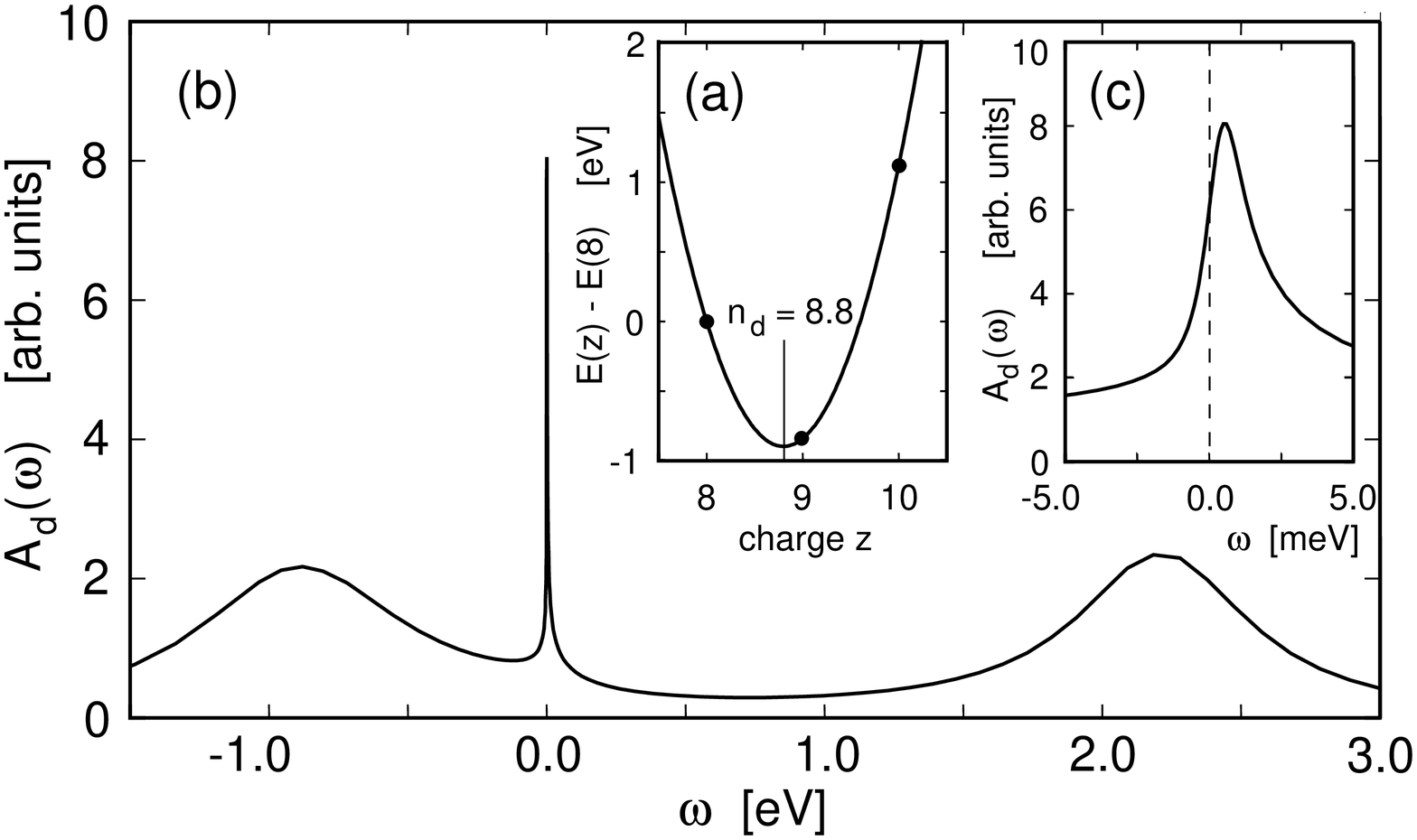,width=8truecm}}
\vspace*{0.5cm}
    \caption{(a) Energy of the 
     relevant charge states $z=8,9,10$ of the Co ion on a Au(111) surface.
     (b), (c) Resulting local d-orbital spectral function $A_d(\omega )$,
     $T=4$K.
     The broad peaks near $\overline\ep_d$ and $\overline\ep_d+\overline U$ 
     and the shift of the Kondo resonance near $\ep _F$ are clearly visible.}
    \label{fig:2}
\end{figure}

For concreteness we focus on a Co atom on Au and
use the Anderson model \cite{Anderson} with a fivefold orbital
degeneracy of the d-level $\ep _d < 0$, $m=1\dots 5$, 
\begin{eqnarray}
  H &=& H_o+
 \ep_{d}\sum\limits_{m,\sigma} d^\dagger_{m,\sigma}
                               d^{\phantom{\dagger}}_{m,\sigma}
  + V\sum\limits_{m\sigma\k} 
    \left( d^\dagger_{m,\sigma} a^{\phantom{\dagger}}_{\k,\sigma} +
    \text{h.c.}\right) \nonumber\\
  &+& U \sum\limits_{(m,\sigma)\neq(m',\sigma ')} 
        d^\dagger_{m,\sigma}d^{\phantom{\dagger}}_{m,\sigma}
        d^\dagger_{m ',\sigma '}d^{\phantom{\dagger}}_{m ',\sigma '}
\label{H0}
\end{eqnarray}
where $H_o=\sum _\k \ep_\k a^\dagger_{\k,\sigma} a_{\k,\sigma}$
is the kinetic energy of the conduction electrons, and 
$a_{\k,\sigma}$, $d_{m,\sigma}$ are the electron operators in the
conduction band and in the d-level, respectively.
$U$ is the Coulomb repulsion between two electrons in any of the local
levels. 
There are two types of resonances in the DOS of the impurity d-level:
($i$) the Co d-levels with a broadening $\Delta =\pi |V|^2\rho_0$ and ($ii$)
the Kondo resonance \cite{Gruner1} whose width is given by the
Kondo temperature $T_K$. 
In order to make contact with experiment, one must obtain realistic
estimates for the parameters of the model. Therefore, we have
applied the semi-relativistic, screened Korringa-Kohn-Rostoker method
\cite{skkr} in combination with the local spin-density approximation 
(LSDA) \cite{lsda}
for calculating the self-consistent electronic structure of a Co 
impurity placed onto a
Au(111) surface. The LSDA assumes a spontaneous magnetization and
consecutive Stoner-like 
splitting of the local level, which is fictitious 
for a single impurity.  
Although this does not account 
for properties of dynamical origin like the splitting into lower and
upper Hubbard states or the Kondo effect, it is an accurate method
\noindent
to determine static quantities like the average 
impurity occupation number $n_d=\sum _{m\sigma}
\langle  d^\dagger_{m,\sigma}d^{\phantom{\dagger}}_{m,\sigma} \rangle $.
This is because these are determined on time scales 
much shorter than the spin flip time $\tau = h /k_B T_K$, 
so that the local moment is effectively static for this purpose.
Our LSDA results are summarized in Fig.~\ref{fig:szk}: The orbital
degeneracy is lifted due to crystal field splitting, and
$\Delta \simeq 0.2$eV. The on-site
Coulomb repulsion $U$ is proportional to the LSDA Stoner splitting
and may be estimated as $U=2.8$eV  \cite{Solovyev}. 
Mainly due to sp-d hybridization,
the Co d-levels are shifted downward compared to bulk Co, so that 
$n_d \simeq 8.8$ instead of $n_d = 7$ expected from the nuclear charge of
Co. The excess charge is compensated by a positive conduction electron
depletion cloud around the impurity. 
It follows that of the 11 possible charge states of the
Co d-level, $z=0,1,\dots ,10$, the system fluctuates only between
$z=8,9,10$. Using Hund's rule, these may be identified with the 
empty ($z=8$, $\sigma=1$, fixed), singly ($z=9$, $\sigma =\pm 1/2$) and
doubly ($z=10$, $\sigma=0$) occupied states of an 
effective spin-1/2 Anderson impurity model without orbital degeneracy.
The level energy $\overline \ep_d$ and Coulomb repulsion $\overline U$
of the effective model are different from those of the original model,
Eq.~(\ref{H0}). They may be extracted as follows \cite{note}:
According to Eq.~(\ref{H0}) and neglecting fluctuations, the charge
states $z=0,1,\dots ,10$ of the d-level have energies
$E(z) = z\ep _d +Uz(z-1)/2$ (Fig.~\ref{fig:2}(a)). 
Since $E(z)$ must have its minimum
(ground state) at $z=n_d\simeq 8.8$, we have $\ep_d/U=-(n_d-1/2)$.
The energies of the singly and doubly occupied orbital, measured relative
to the empty state, $E(8)$, are given by  
$\overline \ep_d=E(9)-E(8)$, $2\overline \ep_d+\overline U=E(10)-E(8)$ 
and may be estimated as 
$\overline \ep_d=-0.84$eV, $\overline U=2.84$eV.
The corresponding Co d-orbital spectral function $A_d(\omega )$, 
calculated using the non-crossing approximation (NCA) \cite{Bickers,Kroha1},
is shown in Fig.~\ref{fig:2} (b). In addition to the broad peaks located near
$\overline \ep_d$ and $\overline \ep_d+\overline U$ the narrow Kondo resonance 
is clearly seen. The Kondo temperature,
$T_K = D \sqrt{2\Delta /(\pi D)}\;{\rm exp}\{ -1/(2\rho_0J) \}$,
with the spin exchange coupling $J = \Delta /(\pi\rho_0)[ 1/|\overline \ep_d| + 
1/|\overline \ep_d+\overline U| ]$, may be estimated as 
$T_K \simeq 52K$. Here $D$ is a band cutoff provided by 
the bulk and (111) surface \cite{Chen} bands of Au, respectively;
we assume $D\simeq\ep _F^{Au} =5.5$eV. 
It follows from the parabolic form of $E(z)$ 
(Fig.~\ref{fig:2} (a)) that $|\overline \ep_d| < (>)\;
\overline\ep_d+\overline U_d$, if $n_d < (>)\; 9$. 
As a consequence of level repulsion, the Kondo resonance is
somewhat shifted upward (present case, $n_d=8.8$;
shown in Fig.~\ref{fig:2} (c)) or downward from the Fermi level. 
According to local Fermi liquid theory \cite{Nozieres}, for $T<T_K$
the Kondo resonance is a pure potential scatterer. For the analytical
treatment below we may, therefore,
model $A_d(\omega )=\frac{1}{\pi}\Im G_d(\omega -i\delta)$ crudely as a 
sum of three Lorentzians. This corresponds to
\begin{eqnarray}
  \label{eq:tomega}
  G_d(\omega -i\delta)&=&\frac {Z_{d}}{\omega - \overline \ep_d -i\Delta}+ 
                         \frac {Z_{U}}{\omega -\overline \ep_d -\overline U 
                                       - i\Delta}
  \nonumber\\ 
 &+&\frac {Z_K}{\omega - \ep_K - iT_K}\ ,
\end{eqnarray}
where $\ep_K$ is the position of the Kondo resonance.
$Z_{d}$, $Z_{U}$ and $Z_K$ are the appropriate strengths of the poles.
In the Kondo regime, at low temperatures ($T\ll T_K$), 
unitary scattering \cite{Nozieres} implies
$Z_K \stackrel{(<)}{\sim} \pi T_K/\Delta $.

We now turn to the conduction electron LDOS as measured by the STM tip at a 
distance $R$ from the impurity. It is related to the electronic field
operator smeared around the tip position {\R}
as $\Psi_{\bf R} = \int \Psi({\bf x}) U_{\bf R}
({\bf x}) \,d^2x$, where $ U_{\bf R}({\bf x})$ is a form factor centered
at ${\bf x}= {\bf R}$, and $\Psi({\bf x})$ is the two-dimensional
Fourier transform of $a_{\k,\sigma}$. According to the Anderson model,
the exact conduction electron $t$-matrix is given in terms of the
d-electron Green's function as $t_{\sigma}(i\omega_n)=
\frac{\Delta}{\pi\rho_0}G_{d,\sigma}(i\omega_n)$. 
Then the correction to the conduction electron 
Green's function due to the presence of the impurity,
$\delta {\cal G}_{R,\sigma}(i\omega_n) = {\cal G}_{R,\sigma}(i\omega_n)
  - {\cal G}^{(0)}_{R,\sigma}(i\omega_n)$, is
\begin{equation}
  \delta {\cal G}_{R,\sigma}(i\omega_n) =  
  {\cal G}^{(0)}_{R,\sigma}(i\omega_n) \,
    t_{\sigma}(i\omega_n) \, {\cal G}^{(0)}_{R,\sigma}(i\omega_n),
\label{deltaG}
\end{equation}
where
${\cal G}^{(0)}_{R,\sigma}(i\omega_n) = -\langle T_\tau
  \bigl (\Psi_{\sigma}(0)\Psi^{\dag}_{R,\sigma} \bigr )\rangle_{\omega_n}$.
The measured perturbation in the tunneling LDOS at distance $R$ is
\begin{equation}
  \delta\rho_R(\omega) = \frac1\pi\Im \delta {\cal
  G}_{R,\sigma}(\omega-i\delta). 
\end{equation}
Using Eq.~({\ref{deltaG}), it can be expressed as
\begin{eqnarray}
  \label{Fanoline}
\delta\rho_{R,\sigma}(\omega) &=& 
\frac{1}{\pi} \bigl[\Im {\cal G}^{(0)}_{R,\sigma}(\omega -i\delta) \bigr]^2 
\times \\
&\phantom{+}&\hspace*{-1.5cm}
\Bigl[\bigl(q_{R,\sigma}^2-1\bigr) \Im t_{\sigma}(\omega-i\delta)
+2q_{R,\sigma} \Re t_{\sigma}(\omega-i\delta) \Bigr]\ ,  \nonumber
\end{eqnarray}
where we have defined 
\begin{equation}
q_{R,\sigma} = 
\frac{\Re {\cal G}^{(0)}_{R,\sigma}(\omega -i\delta)}
     {\Im {\cal G}^{(0)}_{R,\sigma}(\omega -i\delta)}\ .
\label{qfactor}
\end{equation}
In the following we drop the spin index $\sigma$. 
$\Im {\cal G}^{(0)}_{R}(\omega -i\delta)$ and
$q_{R}$ depend on $R$, but on the scale $T_K$ 
very weakly on the energy; thus from here on
$\omega=0$ is taken in these quantities.
At $R = 0$, $q_{R =0}$ is identical to the asymmetry parameter $q$ of
the Fano theory \cite{Fano}, and if $t(\omega -i\delta)$ is a
simple (single-particle) level, $t(\omega -i\delta) = (\Delta /\pi
\rho _0 )/(\omega -\ep_d-i\Delta)$, the line shape of Eq.~(\ref{Fanoline})  
reduces to Fano's well-known expression
$\rho (\omega ) = \rho _0 + \delta\rho
= \rho_0 (x+q)^2/(x^2+1)$, with $x=(\omega -\ep_d)/\Delta$ \cite{Fano}.
However, Eq.~(\ref{Fanoline}) is not limited to the non-interacting case. 
E.g. in the Kondo problem, correlations are contained 
in $t(i\omega _n)$
and may be treated separately (see above), while the Fano line shape is
due to mixing of real and imaginary parts in Eq.~(\ref{Fanoline}).  
The line shape is sensitive both to the scattering phase shift contained
in $t(\omega -i\delta)$ and to the space dependent
phase of the free conduction electron wave function exhibited by $q_{R}$.
In particular, it depends crucially on the local charge density and on
the details of the band structure \cite{Schiller}. 
In the Kondo regime ($|\overline \ep_d|,\; \overline U \gg \Delta$,
$T\ll T_K$), for $\omega \stackrel{<}{\sim} T_K$, $|\ep_K| < T_K\ll \Delta$,
and using the (approximate) unitarity relation
$Z_K \stackrel{(<)}{\sim} \pi T_K/\Delta$ (see remark after
Eq.~(\ref{eq:tomega})) the $t$-matrix may be cast in the form
\begin{mathletters}
\begin{equation}
  \Re t(\omega-i\delta)=\frac {1}{\rho_0}\biggl \{ 
  \frac {\ep}{1+\ep^2} + \beta + 
  {\cal O}(\frac{T_K}{\Delta})\biggr
  \}
\end{equation}
\begin{equation}
  \Im t(\omega-i\delta)=\frac {1}{\rho_0}\biggl \{ 
  \frac 1{1+\ep^2} 
  - \beta\frac{\Delta}{\overline\ep_{d}} + 
{\cal O}(\frac{T_K}{\Delta})\biggr \}\ ,
\end{equation}
\end{mathletters}
where
$\beta=-(Z_d/\pi) \;
    \overline\ep_{d}\Delta/({\overline\ep_{d}^2 +\Delta^2})$,
and $\ep=(\omega-\ep_K)/{T_K}$. In $\beta$ we have taken into account
only the 9-fold occupied d-state, as this is closest to $\ep _F$ and
the other charge states give only a small contribution to the
frequency independent part of $t(\omega )$.
Thus the final expression for the LDOS correction is
\begin{equation}
  \delta\rho_{R}(\omega) = 
   \frac{[\Im {\cal G}^{(0)}_{R,\sigma}(\omega -i\delta)]^2}{\pi\rho _0}\,
  \biggl \{
  \frac{2q_{R}\ep + q_{R}^2-1}{\ep^2+1}+ C_R \biggr \} \ , 
\label{roveg}
\end{equation}
where $C_R= \beta [2q_{R} - (q_{R}^2-1)
{\Delta}/{\overline\ep_{d}}]$
arises from potential scattering by the
d-level and corresponds to a weakly energy dependent
Friedel oscillation. The first part coming
from the scattering by the Kondo resonance gives a Fano line
shape in the tunneling LDOS, controlled by the parameter $q_{R}$.

Eq.~(\ref{roveg}) can be fitted to the experimental 
data for a Co atom on a
Au (111) surface \cite{Madhavan} with excellent agreement 
(Fig.~\ref{fig:3}).
From the fit parameters (see the figure caption of
Fig.~\ref{fig:3}) we can conclude that $\ep_K>0$.  The
value of the Kondo temperature $T_K\simeq 50$K obtained from the fit is
substantially smaller than the bulk value ($T_K>300$ K for Co
\cite{Gruner1}) as the coupling of the impurity to the conduction
electrons is weaker on the surface. Both $T_K$ and the shift $\ep_K$
of the Kondo resonance are consistent with the predictions
of the NCA calculation 
in combination with the LSDA for Co on a Au(111) surface.
\begin{figure}
\centerline{\psfig{figure=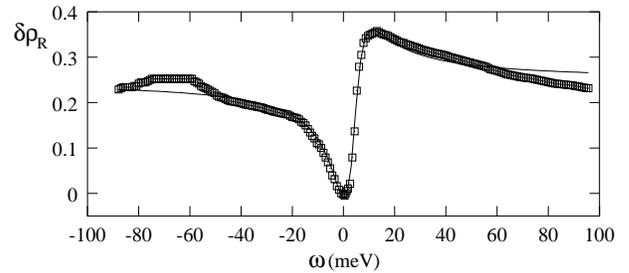,width=8truecm}}
\vspace*{0.5cm}
    \caption{Fit of Eq.~(8) (solid line) to the experimental data 
      [2] (squares) at $R=0$. The fit parameters are
      $q_{R=0} =0.66$, $C_{R=0}= 0.95$ and
      $\ep_K=3.6$meV,  $T_K=5$meV$\sim 50$K.}
\label{fig:3}
\end{figure}

The remaining task is to calculate $q_{R}$ and $C_{R}$ as a function of
$R$. We have considered tunneling of electrons from the tip (1)
into the 3-dimensional Au bulk states as well as (2) into the 2-dimensional 
Au(111) surface band \cite{Chen}. In both cases a free electron-like
band structure was assumed, neglecting, e.g., in case (2) corrections due to the 
herringbone surface reconstruction \cite{Chen}. 
In case (1) the bulk Au Fermi wave number was
taken as $k^{(b)}_F=1.21 \AA ^{-1}$, while in case (2) it was extracted from
the surface band structure mapped out in Ref.~\cite{Chen} by scanning
tunneling spectroscopy, $k^{(s)}_F = 0.189 \AA ^{-1}$. 
A window function was adopted for the form factor $U_{\R}(\x )$,
and the scattering by the Co d-orbital was taken to be isotropic, 
as the scattering phase shift does not strongly
depend on angular momentum. 
From Eq.~(\ref{roveg}), the Lorentzian line shape 
with a weight $A = \Re[{\cal G}_{R}^{(0)}]^2/(\pi\rho_0)$ 
is formed at those distances $R$ where $\Re{\cal G}_{R}^{(0)}=0$ ($A<0$) or
$\Im{\cal G}_{R}^{(0)}=0$ ($A>0$). For case (1), the R dependence
of the LDOS is demonstrated in Fig.~\ref{fig:4}. The line shape changes
periodically between asymmetric Fano and Lorentzian line shape, 
the period given by the Friedel wave length, 
$\lambda ^{(b)}_F = \pi/k^{(b)}_F\simeq 2.6\AA$.
The overall amplitude decreases with increasing distance. 
For case (2), the results are similar, however with an oscillation period of
$\lambda ^{(s)}_F = \pi/k^{(s)}_F\simeq 16.6\AA$. The latter is in
good quantitative agreement with the measured \cite{Madhavan} 
period of approximately $16\AA$, indicating that the two-dimensional
Au(111) surface band gives an important contribution to the total
LDOS measured by STM. However, the precise dependence of
the line shape on $R$ is not reproduced by our
simplifying assumption of a free electron band structure. 
In particular, the model calculation predicts an 
initially increasing $q_{R}$ as $R$ grows from $R=0$, 
while the fit of Eq.~(\ref{roveg}) to the 
experiments \cite{Madhavan} implies a decreasing $q_{R}$.
The precise $R$-dependence of $q_{R}$ will require taking into account the 
detailed band structure as well as the additional scattering phase shift
induced by the charge of the Co ion and its spatially extended screening
cloud. This is beyond the scope of the present Letter.
\begin{figure}
\centerline{\psfig{figure=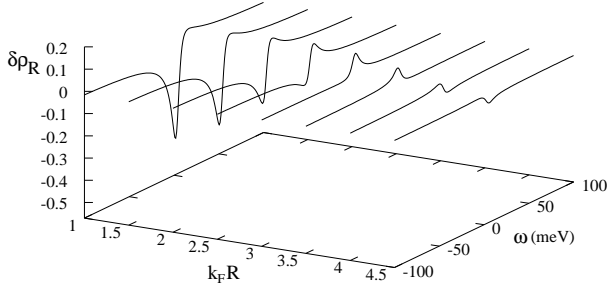,width=8truecm}}
\vspace*{0.5cm}
    \caption{Qualitative dependence of the line shape of the 
      tunneling DOS on the distance of the tip from the
      impurity using the first part of Eq.~(\ref{roveg}).}
    \label{fig:4}
\end{figure}
 
Summarizing, we have shown that the Kondo resonance
in the magnetic impurity d-level density of states ($T<T_K$) causes a
Fano line shape in the density of states measured by STM even if
we take into account only the tunneling into the conduction electron
states. As expected, the Fano shape arises independently of whether the
local level is a single-particle orbital or a many-body resonance
like in the Kondo effect.
Superimposed on the Fano line is a weakly energy dependent Friedel oscillation
induced by potential scattering off the broad d-level. 
From the fit to the experimental data \cite{Madhavan} for Co on Au(111)
surfaces we conclude that
the Kondo resonance is shifted upward from the Fermi level, and
$T_K$ for Co on the surface is substantially
reduced compared to the bulk value. By combining electronic structure
calculations (LSDA) with methods for strong correlations (NCA) these
findings are reproduced semi-quantitatively. In particular, the shift
of the Kondo resonance is due to level repulsion between the 
d-level and the Kondo resonance.  
We demonstrated the dependence of the line shape on the
distance of the tip from the impurity by using bulk as well as surface state 
Green's functions. 
We emphasize, however, that details of the band structure have to be 
taken into account in order to reproduce this distance dependence 
quantitatively.\\
\indent
We acknowledge useful discussions with R. Berndt, M. F. Crommie and 
W.-D.~Schneider. 
J.K. is grateful for the hospitality of the Condensed Matter Physics group,
Technical University of Budapest, where part of this work was performed.
This work was supported by the
Humboldt foundation (A.Z.), by the OTKA Postdoctoral Fellowship D32819 (O.\'U.),
by Hungarian grants
OTKA T024005, T029813, T030240 and by DFG through SFB 195 (J.K.).

}
\end{multicols}


\begin{thebibliography}{99}
\bibitem{note} 
A detailed derivation will be published elsewhere.
\end{thebibliography}
\end{document}